\documentclass{emulateapj}

\usepackage{graphicx}

\newcommand{\Mstar}{\ensuremath{M_{\ast}}}
\newcommand{\Msun}{\ensuremath{M_{\odot}}}

\newcommand{\Rstar}{\ensuremath{R_{\ast}}}
\newcommand{\Rkep}{\ensuremath{R_{\rm K}}}
\newcommand{\Ralf}{\ensuremath{R_{\rm A}}}
\newcommand{\Resc}{\ensuremath{R_{\rm E}}}
\newcommand{\Rsun}{\ensuremath{R_{\odot}}}

\newcommand{\estar}{\ensuremath{\eta_{\ast}}}
\newcommand{\Bstar}{\ensuremath{B_{\ast}}}
\newcommand{\Mdot}{\ensuremath{\dot{M}}}

\newcommand{\Vinf}{\ensuremath{V_{\infty}}}
\newcommand{\Veq}{\ensuremath{V_{\rm eq}}}
\newcommand{\Vcrit}{\ensuremath{V_{\rm crit}}}

\newcommand{\vphi}{\ensuremath{v_{\phi}}}
\newcommand{\vkep}{\ensuremath{v_{\rm K}}}
\newcommand{\vesc}{\ensuremath{v_{\rm E}}}

\newcommand{\gauss}{\ensuremath{\rm G}}

\newcommand{\kelv}{\ensuremath{\rm K}}
\newcommand{\ksecs}{\ensuremath{\rm ks}}
\newcommand{\years}{\ensuremath{\rm yr}}
\newcommand{\keV}{\ensuremath{\rm keV}}

\newcommand{\sOriE}{$\sigma$~Ori~E}
\newcommand{\tOriC}{$\theta^{1}$~Ori~C}
\newcommand{\tOriA}{$\theta^{2}$~Ori~A}

\begin{document}

\title{Centrifugal Breakout of Magnetically Confined Line-Driven Stellar Winds}  
\shorttitle{Centrifugal Breakout of Magnetic Stellar Winds}  

\author{Asif ud-Doula$^{\ast}$, Richard H.~D. Townsend, and Stanley P. Owocki}  

\affil{Bartol Research Institute, Department of Physics \& Astronomy,
       University of Delaware, Newark, DE 19716}

\altaffiltext{$\ast$}{also at Department of Physics, Swarthmore College, Swarthmore, PA 19081}

\begin{abstract}
We present 2D MHD simulations of the radiatively driven outflow from a
rotating hot star with a dipole magnetic field aligned with the star's
rotation axis. We focus primarily on a model with moderately rapid
rotation (half the critical value), and also a large magnetic
confinement parameter, $\estar \equiv \Bstar^2 \Rstar^{2}/ \Mdot \Vinf
= 600$. The magnetic field channels and torques the wind outflow into
an equatorial, rigidly rotating disk extending from near the Kepler
corotation radius outwards.  Even with fine-tuning at lower magnetic
confinement, none of the MHD models produce a stable Keplerian disk.
Instead, material below the Kepler radius falls back on to the stellar
surface, while the strong centrifugal force on material beyond the
corotation escape radius stretches the magnetic loops outwards,
leading to episodic breakout of mass when the field reconnects.  The
associated dissipation of magnetic energy heats material to
temperatures of nearly $10^{8}\,{\rm K}$, high enough to emit hard
(several keV) X-rays.  Such \emph{centrifugal mass ejection}
represents a novel mechanism for driving magnetic reconnection, and
seems a very promising basis for modeling X-ray flares recently
observed in rotating magnetic Bp stars like $\sigma$~Ori~E.
\end{abstract}

\keywords{MHD --- stars: winds --- stars: magnetic fields --- stars:
rotation --- stars: flare --- X-rays: stars}

\section{Introduction} \label{sec:intro}

Magnetic fields can greatly influence stellar winds by guiding the
outflowing plasma along field lines
\citep[e.g.,][]{MacG1988,ShoBro1990}.  The extent of this influence
depends largely on the magnetic field strength and the wind mass loss
rate.  Two-dimensional (2D) magnetohydrodynamical (MHD) simulations of
axisymmetric hot-star winds with no rotation show that the competition
between the wind and the field can be quite well characterized by a
single wind magnetic confinement parameter, \estar, as defined by
\citet{udDOwo2002}, and described further in \S\ref{sec:alfven} below.

For significant confinement (i.e., $\estar \ga 10$), material driven
from opposite footpoints of closed magnetic loops is forced to collide
near the loop tops, leading to a \emph{Magnetically Confined Wind
Shock} (MCWS) scenario for production of intermediate-hardness X-rays
\citep{BabMon1997a,BabMon1997b}.  For the magnetic hot star \tOriC,
which is inferred to have both a moderately strong wind (with mass
loss rate $\approx 10^{-7}$\,\Msun/\years) and a large-scale, dipole
field of strength $\Bstar \approx 1100\,\gauss$ \citep{Don2002},
recent MHD simulations based on this MCWS paradigm provide quite good
agreement with the X-ray spectrum observed by \emph{Chandra}
\citep{Gag2005}.  Indeed, because \tOriC\ is a relatively slow rotator
(with period $\sim$~15~days), even the observed rotational modulation
can be well accounted for by simply varying the observational
perspective within a {\em non-rotating} dynamical simulation model.

The purpose of this letter is to report on initial MHD simulation
results for more rapidly rotating hot stars, in which the rotation can
have a direct dynamical effect on the overall evolution of the wind
outflow and its confinement.  The results have particular relevance in
the context of two proposed models for rotating, magnetic hot stars,
known generally as the \emph{Magnetically Torqued Disk} (MTD) model
\citep{Cas2002} and the \emph{Rigidly Rotating Magnetosphere} (RRM)
model \citep{TowOwo2005}.  The MTD has been proposed as a scenario for
producing \emph{Keplerian} decretion disks, such as inferred from the
Balmer line emission of classical Be stars.  The RRM leads instead to
\emph{rigid-body} disks or clouds that provide a promising way to
explain the rotationally modulated Balmer emission from strongly
magnetic Ap/Bp stars like the B2pV star \sOriE\ \citep{Tow2005}.

After first casting the relative importance of magnetic confinement
vs.\ rotation in terms of characteristic Alfv{\`e}n vs.\ Kepler
corotation radii (\S\ref{sec:alfven}), we show in
\S\ref{sec:keplerian} that even for parameters fine-tuned to optimize
the chances of a producing a Keplerian disk, our 2D MHD simulations
for rotationally aligned, dipole fields instead produce a combination
of inflow and outflow for material below and above the Kepler
corotation radius, and never lead to the kind of extended Keplerian
disk envisioned in the MTD scenario.

On the other hand, we show in \S\ref{sec:breakout} that a model with
strong magnetic confinement ($\estar = 600$) does lead to a
\emph{rigid-body} disk, much as predicted in the RRM model.  However,
the outer regions of this disk are characterized by episodic breakout
of material, driven by the strong net outward centrifugal force
associated with the spin-up of material into corotation with the
underlying star.  Moreover, the energy release associated with
magnetic reconnection in these breakout events heats material to
extremely high temperatures, providing a promising mechanism for
explaining the very hard X-ray flare emission recently observed from
magnetic Bp stars such as \sOriE.  We conclude in \S\ref{sec:summary}
with a brief summary of our findings, noting also the need for further
work on the detailed physics of this somewhat-novel, centrifugally
driven reconnection mechanism.

\section{Alfv{\`e}n vs.\ Kepler Corotation Radii} \label{sec:alfven}

MHD simulations \citep{udDOwo2002,udD2003} indicate that the
effectiveness of magnetic fields in channeling a stellar wind outflow
can be characterized by the ratio between the magnetic and wind energy
densities
\begin{equation} \label{eqn:etadef}
 \eta(r) \equiv \frac{B^2/8 \pi}{\rho v^{2} / 2}
              = \estar \frac{(r/\Rstar)^{2-2q}}{(1-\Rstar/r)^{\beta}}.
\end{equation}
This ratio furnishes an indication of whether the magnetic field at
radius $r$ dominates the wind outflow ($\eta \gg 1$), or vice versa
($\eta \ll 1$). As indicated by the second equality, its overall
scaling is set by the ``magnetic confinement parameter'' $\estar
\equiv \Bstar^{2} \Rstar^{2} / \Mdot \Vinf$, which depends on the
strength of the field \Bstar\ at the stellar surface radius
$r=\Rstar$, and on the wind terminal momentum $\Mdot \Vinf$. The
radial variation of $\eta$ is then modeled in terms of a magnetic
power-law index $q$ ($=3$ for a dipole) and a velocity index $\beta$
($ \approx 1$ for a standard CAK wind).  If for simplicity we ignore
the wind velocity variation (i.e. by taking $\beta=0$), the Alfv{\`e}n
radius -- at which the magnetic and wind energy densities are equal,
$\eta(\Ralf) \equiv 1$ -- can be solved explicitly as
\begin{equation} \label{eqn:Ralf}
 \Ralf = \estar^{1/4} \Rstar.
\end{equation}
As shown by simulation results summarized in \citet{udDOwo2002}, this
Alfv{\`e}n radius provides a reasonable estimate for the maximum
radius of magnetic loops that remain closed in a wind outflow.
Moreover, since in rotating models such closed loops tend to keep the
outflow in rigid-body rotation with the underlying star, it also
defines the radius of maximum rotational spin-up of the wind azimuthal
velocity \vphi.

To characterize such rotational effects, let us next define a Kepler
corotation radius \Rkep. Analogous to the terrestrial geostationary
orbital radius, \Rkep\ is the point at which the centrifugal force
arising due to corotation balances the local gravitational force from
the star. That is,
\begin{equation}
 \frac{G\Mstar}{\Rkep^{2}} =
 \frac{\vphi^{2}}{\Rkep} =
 \frac{\Veq^{2}}{\Rstar^{2}},
\end{equation}
where \Veq\ is the equatorial rotation velocity of the star. Solving
this equation, we find the Kepler radius as
\begin{equation} \label{eqn:Rkep}
 \Rkep = W^{-2/3} \Rstar,
\end{equation}
where $W \equiv \Veq/\Vcrit$ is ratio of the star's equatorial
velocity to the critical velocity $\Vcrit \equiv
\sqrt{G\Mstar/\Rstar}$.

Finally, it is also worth noting here that corotation out to an only
slightly higher escape radius,
\begin{equation} \label{eqn:Resc}
 \Resc = 2^{1/3} \Rkep = 2^{1/3} W^{-2/3} \Rstar,
\end{equation}
would imply an azimuthal velocity \vphi\ that equals the local escape
speed from the star's gravitational field.

\section{Attempts to Fine Tune for a Keplerian Disk} \label{sec:keplerian}

\begin{figure}
\begin{center}
\plotone{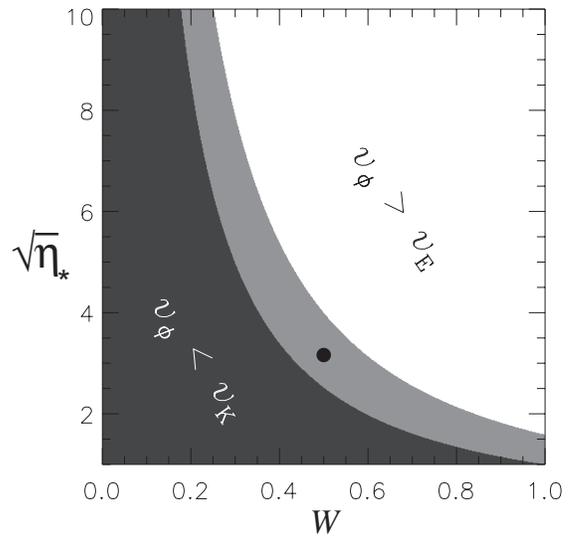}
\caption{The key domains in a parameter plane of magnetic field
strength (as represented by $\sqrt{\estar} \sim \Bstar$) vs.\ stellar
rotation (as represented by the critical rotation fraction $W \sim
\Veq$).  The black area to the lower left denotes models in which the
magnetic spin-up of the azimuthal velocity \vphi\ is insufficient to
reach the velocity \vkep\ required for Keplerian orbit, while the
white area in the upper right denotes models in which the spin-up
leads to \vphi\ that exceed the escape speed \vesc. Only in the gray
area, bounded between curves for models defined by $\Ralf = \Rkep$ and
$\Ralf = \Resc$, does spin-up into Keplerian orbit seem possible. The
dot positioned in the middle of this area indicates the parameters of
the MHD simulations described in \S\ref{sec:keplerian}.}
\label{fig:domains}
\vspace{-0.2cm}
\end{center}
\end{figure}

The above scalings suggest that a likely necessary condition for
propelling outflowing material into a Keplerian disk (as envisioned in
the MTD scenario for classical Be stars) is to choose a combination of
parameters for magnetic confinement vs.\ stellar rotation such that
$\Rkep < \Ralf < \Resc$.  In the parameter plane of $\sqrt{\estar}
\sim \Bstar$ vs.\ $W \sim \Veq$ illustrated in Fig.~\ref{fig:domains},
the required combination is represented by the relatively narrow gray
band.

To obtain a model that has the best chance of yielding a Keplerian
disk, we therefore conduct 2D axisymmetric MHD simulations of a star
having parameters $\estar = 10$ and $W = 0.5$ that fall in the middle
of the gray area in Fig.~\ref{fig:domains}. The initial magnetic field
geometry is assumed to be a dipole aligned with the rotation
axis. Calculations are undertaken using the publicly available ZEUS-3D
code \citep{StoNor1992}, augmented to include radiative driving and
radiative cooling \citep{Gag2005}, with typical early B-star values
for the stellar radius and mass, $\Rstar = 6.4\,\Rsun$ and $\Mstar =
11\,\Msun$.  The numerical grid and boundary conditions follow those
given in \citet{udDOwo2002}. Since the star is rotating at one half
its critical value, the oblateness and gravity darkening of the
stellar surface are relatively small, and are neglected in our
simulations.

\begin{figure}
\plotone{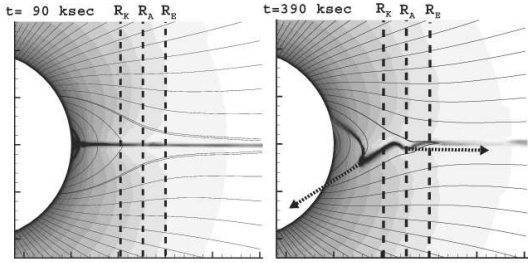}
\caption{Density in a 2D MHD simulation with $\estar = 10$ and $W =
1/2$, shown at time snapshots of 90\,\ksecs\ (left) and 390\,\ksecs\
(right) after a dipole field is introduced into an initially steady,
spherical, unmagnetized, line-driven stellar wind.  The curves denote
magnetic field lines, and the vertical dashed lines indicate the
equatorial location of the Kepler (\Rkep), Alfv{\`e}n (\Ralf), and
escape (\Resc) radii defined in eqns.~(\ref{eqn:Rkep}),
(\ref{eqn:Ralf}), and (\ref{eqn:Resc}), respectively The arrows denote
the upward and downward flow above and below the Kepler radius,
emphasizing that the material never forms a stable, orbiting disk.}
\label{fig:fallback}
\vspace{-0.1cm}
\end{figure}

Figure~\ref{fig:fallback} illustrates results from the simulation.
The left-hand panel shows the logarithm of density (superimposed with
field lines) at a time 90\,\ksecs\ after the introduction of the
dipole field into a spherically symmetric wind.  Superficially, these
conditions do temporarily resemble the MTD scenario.  However, a
closer inspection shows that most of the density-enhanced equatorial
region does not have the appropriate \vphi\ necessary for a stable,
stationary, Keplerian orbit.  Indeed, in just a few \ksecs\ of
subsequent evolution this putative ``disk" becomes completely
disrupted, characterized generally by infall of the material in the
inner region, i.e. below the Kepler radius \Rkep, and by outflow in
the outer region above the Kepler radius. The right-hand panel
demonstrates this by showing the irregular form of the dense
compression at an arbitrarily chosen later time (390\,\ksecs).  The
bold arrows emphasize the flow divergence of the dense material both
downward and upward from the Kepler radius.  (The overall evolution is
most vividly illustrated through animations available on the
internet\footnote{\tt
http://www.bartol.udel.edu/\~{}owocki/animations/den4wp5eta10.avi}.)

We have carried out a number of other similar MHD simulations
exploring parameter space in both the rotation rate and magnetic field
strength directions.  In every case, we find that the equatorial
compressions are dominated by similar radial inflows and/or outflows,
with no stable Keplerian disks forming.  We thus conclude that
stationary, large-scale magnetic fields are dynamically ill-suited for
producing Keplerian or near-Keplerian decretion disks.

\section{A Centrifugal Breakout Mechanism For X-Ray Flaring} \label{sec:breakout}

A recent study by \citet{TowOwo2005} suggests that, in the limit of
very strong magnetic confinement ($\estar \gg 1$), magnetic torquing
can readily produce {\it rigid-body} disks.  In this strong-field
limit, the field lines can be idealized as rigid pipes that channel
the wind outflow and force it to corotate with the underlying star.
This leads to a rigidly rotating magnetosphere in which material
accumulates near minima of the effective gravito-centrifugal potential
along each field line.  \citet{Tow2005} show that the resulting
circumstellar clouds provide a natural explanation for the observed
rotational modulation of both the photometric continuum and Balmer
line emission in the B2pV star \sOriE.

A key feature of this RRM model is that the magnetic field not only
torques up outflowing material into corotation around the star, but
then also \emph{holds it down} against the net outward centrifugal
acceleration beyond the Kepler radius.  In the semi-analytic RRM
analysis, the field is idealized as being strong enough to enforce
corotation and confinement at arbitrarily large radii, and for
arbitrarily long times.  More realistically, however, for large but
finite field strengths, the secular accumulation of material into the
RRM means that these idealizations must break down once the total
outward centrifugal force exceeds a level that can be balanced by the
finite magnetic tension.

\citet{TowOwo2005} speculate that this will lead to centrifugal
breakout of material, and give estimates for the expected
characteristic breakout timescale as a function of distance above the
Kepler radius (see their Appendices A and B).  For the case of \sOriE,
they estimate that the largest-scale evacuations can be expected every
$\sim 10-100\,\years$, but they also anticipate a whole hierarchy of
breakout events extending down to timescales of days.

\begin{figure*}
\leavevmode
\begin{centering}
\includegraphics[scale=0.45,angle=270]{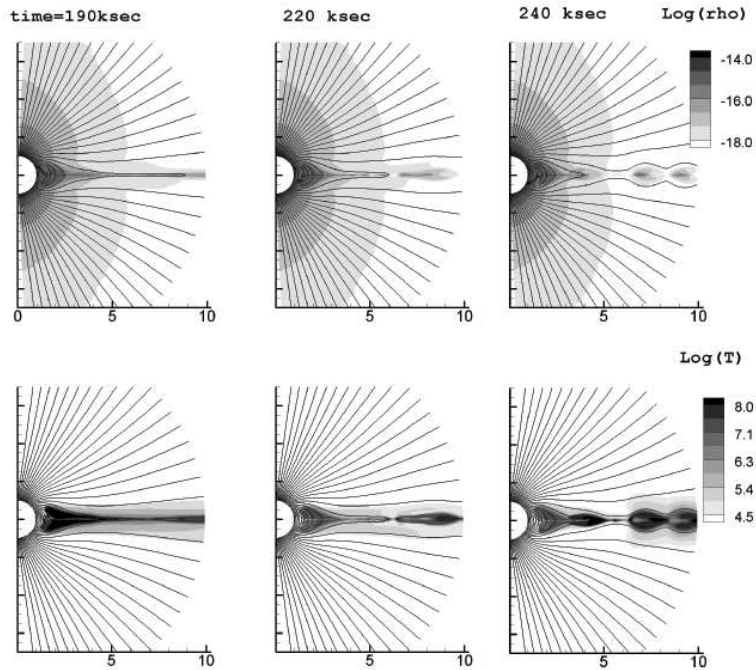}
\caption{MHD simulation results for a radiatively driven stellar wind
from a star rotating at half the critical rate, with a
rotation-aligned dipole field that has a magnetic confinement
parameter $\estar = 600$.  The grayscales show the logarithm of
density (top panels) and temperature (bottom panels) at three sample
time snapshots, chosen to show the centrifugal ejection of material
accumulated in an equatorial disk.  The associated stretching and
eventual reconnection of the field lines heats material to
temperatures of order $10^{8}\,\kelv$, hot enough to emit hard (few
\keV) X-rays.}
\label{fig:breakout}
\end{centering}
\end{figure*}

Our present MHD simulations allow us to examine the dynamics of this
breakout process.  Unfortunately, at the confinement parameter $\estar
\sim 10^{7}$ appropriate to \sOriE, the Alfv{\`e}n speed becomes very
large, implying a very small Courant time step to ensure numerical
stability; this together with the full 3D geometry needed to model the
oblique-field tilt relative to the rotation axis makes direct MHD
simulation of the \sOriE\ case prohibitively expensive.  Instead, we
consider the more tractable, 2D axisymmetric case of a field-aligned
dipole with only moderately strong magnetic confinement ($\estar =
600$).  As appropriate for \sOriE, we again set the rotation to half
the critical rate ($W = 0.5$), with other stellar and grid parameters
also the same as used in \S\ref{sec:keplerian}.

Snapshots from the simulations, plotted in Fig.~\ref{fig:breakout},
reveal that there forms a rigidly rotating disk near and above the
Kepler radius $\Rkep/\Rstar \approx 0.5^{-2/3} \approx 1.6$, as
predicted by the RRM model.  But in the outer regions of this disk
(near the Alfv{\`e}n radius $\Ralf/\Rstar \approx 600^{1/4} \approx
5$), there is indeed also a semi-regular sequence of breakout events.
During a breakout, the magnetic field lines become so drawn out by the
ejected plasma that they reconnect, snapping back toward the star
(cf. middle and right panels of Fig.~\ref{fig:breakout}).  The energy
release associated with this reconnection, and its subsequent
dissipation via radiative cooling, represents a strong candidate for
producing the hard X-ray flares observed in \sOriE\ by \emph{ROSAT}
\citep{GroSch2004} and \emph{XMM-Newton} \citep{San2004}.

This scenario is reminiscent of fast magnetic reconnection processes
invoked for solar flares \citep[e.g.,][]{Pet1966,Nit2004}. But in the
present case, the work to stretch the magnetic fields comes from the
rotational energy of the dense plasma at the edge of the disk, with
the trigger for reconnection being the wind impingent at an oblique
angle from higher latitudes.  The centrifugal energy stored
temporarily in the magnetic field is thereby converted into thermal
energy, heating nearby material to temperatures $T \sim
10^{8}\,\kelv$, high enough to produce the hard ($\geq 2\,\keV$)
components of the X-ray flares observed in \sOriE. A similar mechanism
may also be relevant to X-ray outbursts recently observed in other
(possibly magnetic) early-type stars, such as the mid-B star HD~38563S
\citep{Yan2004} and the late-O star \tOriA\ \citep{Fei2002}.

The centrifugal breakout here provides an interesting counterexample
to the ``slingshot prominence'' model of \citet{JarBal2005}.  This
cool-star, coronal model envisions a tearing-mode instability
\citep{Fur1963} in the current sheet dividing open field regions in an
overlying, gas-pressure-driven wind.  This leads to a magnetic
reconnection that effectively \emph{traps} some of the outflowing wind
material, contributing then to an extended, quasi-static coronal
prominence \citep{Jar2001}.  Although the centrifugal support of this
prominence is similar to our RRM scenario for hot-stars, in this
cool-star case the high coronal back-pressure chokes off any footpoint
upflow that could continuously feed the accumulation region above the
Kepler radius.  In our hot-star model, it is this secular accumulation
of material that eventually forces centrifugal breakout and drives the
associated reconnection.  As such, unlike the tearing-mode
reconnection postulated in this cool-star context, we believe the
overall rate of reconnection and mass ejection should be relatively
independent of the specific instabilities and plasma processes at
play.

In real plasmas, the details for magnetic reconnection can be quite
complex, involving non-thermal particle acceleration \citep{Wei1995}
and/or high-speed jets that form dissipative shocks \citep{Pet1966},
with the distribution of heat modified by thermal conduction or
non-local particle transport \citep{MiyYok2003}. The Zeus-3D numerical
code used here is based on ideal an MHD formalism that effectively
ignores these complexities, and indeed does not even include any
explicit resistivity that could heat material via Ohmic dissipation.
Instead, magnetic reconnection is enabled by the finite spatial grid
resolution, leading to grid-scale currents inducing high-speed jets
that quickly dissipate the flow energy in localized shock fronts.
Nonetheless, the reconnection rates observed in our models are
consistent with kinetic simulations \citep[e.g.,][]{Sha1999}, and the
code still approximately fulfills global energy conservation, thus
providing a first description of the net dissipation of magnetic
energy into plasma heating.  Hence, we believe the simulations
presented here represent a physically reasonable scenario for heating
by centrifugally driven reconnection, with the associated X-ray
emission providing a promising explanation for the X-ray flares
observed from magnetic Ap/Bp stars.

\section{Summary} \label{sec:summary}

The rigid-body corotation makes the equatorial disks in our
simulations quite distinct from the Keplerian disk envisioned in the
MTD scenario promoted by \citet{Cas2002}. We find no evidence of an
extended region of Keplerian rotation.  Instead the results correspond
more closely to those derived in the RRM analysis of
\citet{TowOwo2005}, although the confinement parameter we consider is
near the lower end of the range of validity for the basic rigid-field
approach.

Moreover, our numerical simulations further suggest a novel mechanism
for magnetic reconnection, driven by episodic centrifugal breakout
events.  The strong reconnection heating associated with such breakout
represents a promising mechanism for explaining X-ray flares observed
in magnetic Bp stars such as \sOriE.

In future work, we plan to use more-extensive numerical tools to
investigate in greater detail the reconnection processes within this
centrifugally driven breakout scenario.  We also plan to explore
hybrid methods that could allow extensions of MHD simulations to
larger confinement parameters, including 3D models of tilted-field
cases more appropriate to \sOriE\ and other magnetic hot-stars.

\acknowledgements

We thank Michael Shay and an unnamed referee for their helpful
comments on the physics of reconnection. We acknowledge support from
NASA grants LTSA/NNG05GC36G and \emph{Chandra}/GO3-3024C.

\end{document}